\documentclass[11pt]{article}

\usepackage[a4paper,margin=1in]{geometry}
\usepackage{amsmath,amssymb}
\usepackage{graphicx}
\usepackage{booktabs}
\usepackage{array}
\usepackage{siunitx}
\usepackage{microtype}
\usepackage{caption}
\usepackage{url}
\usepackage[hidelinks]{hyperref}
\usepackage{float}

\title{Artificial Neural Network Assisted Modelling of Tangent Galvanometer Measurements for the Determination of Horizontal Component of Earth's Magnetic Field}
\author{
Saralasrita Mohanty$^{1,*}$, Sudakshina Prusty$^{1,*}$, Anshuman Pal$^{2}$, Pradipta Kumar Mishra$^{2}$\\[3pt]
\small $^{1}$School of Physical Sciences, National Institute of Science Education and Research, Jatni, Odisha 752050, India\\
\small $^{2}$Faculty of Engineering and Technology, Sri Sri University, Cuttack, Odisha 754006, India\\
\small $^{*}$Corresponding authors: \texttt{saralasrita@niser.ac.in}; \texttt{sprusty@niser.ac.in}
}
\date{}

\begin{document}
\maketitle

\begin{abstract}
The Tangent Galvanometer (TG) is a standard undergraduate laboratory experiment for estimating the horizontal component of Earth's magnetic field ($B_H$) by reading the angle of deflection of the magnetic needle corresponding to the current flowing through a circular coil. In this study, an artificial neural network (ANN) is used as a complementary data-driven model to predict the value of $B_H$. A dataset comprising 225 observations obtained with 50- and 500-turn coils is utilized for developing the ANN model. The data are divided into training (70\%), validation (15\%), and testing (15\%) subsets. The model was optimized using three-hidden-layer and Tanh as activation function. Three different models (Models A, B and C) with different number of input parameters are tested for optimum performance in terms of error metrics. Model C uses the highest number of input parameters such as the input current, angle of deflection $\theta$, $\tan\theta$ and induced magnetic field. The first two parameters are taken from direct experimental measurements and are independent parameters. The second two are derived from the first two. By computing the error metrics for each of the models, it is shown that errors are minimized in Model C while predicting the value of $B_H$ as compared to an adopted local geomagnetic horizontal reference value of 39.0~$\mu$T. The results demonstrate the usefulness of ANN as a complementary tool for analyzing experimental variability.
\end{abstract}


\section{Introduction}
The Tangent Galvanometer is a classical undergraduate laboratory arrangement used to study the magnetic field produced by a current-carrying circular coil and its interaction with the horizontal component of Earth's magnetic field. When the coil is oriented in the magnetic meridian, the field at its center is approximately perpendicular to the horizontal geomagnetic field, and the resulting needle deflection illustrates the tangent law~\cite{beggan2026,griffiths2017,squires2001}.

The conventional analysis is physically transparent: the current and deflection angle are measured, the magnetic field produced by the coil is calculated, and the horizontal component of Earth's field, $B_H$, is obtained from the tangent-law relation. In practice, however, the calculated values vary from observation to observation because of instrument resolution, current fluctuations, imperfect alignment, pivot friction, parallax, coil tolerances, and environmental magnetic disturbances.

Artificial neural networks (ANNs) provide flexible nonlinear regression models that can learn relationships between measured or derived variables and an experimental target~\cite{bishop2006,haykin2009,bishop2024,lecun2015,goodfellow2016}. Their use in undergraduate laboratory analysis can provide a direct way of connecting familiar physical measurements with data-driven modelling.

In the present study, current ($I$), deflection angle ($\theta$), and the number of turns ($N$) are used as basic experimental inputs. The analysis is then extended by adding two physically derived representations: $\tan\theta$ and the magnetic field produced at the coil centre, $B_i$. The study also examines network depth and activation function, the effect of feature representation, prediction residuals, performance for the two coil configurations, and comparison with simpler regression baselines. The resulting predictions are compared with the conventional experimental estimates and with an adopted local geomagnetic horizontal reference. This provides a computational extension of a standard laboratory experiment while keeping the physical meaning of the variables explicit.

\section{Tangent Galvanometer: Physical Principle}
\subsection{Magnetic field produced by the coil}
For a circular coil of radius $R$, carrying current $I$ and having $N$ turns, the magnetic field at the centre of the coil is
\begin{equation}
B_i = \frac{\mu_0 N I}{2R},
\label{eq:bi}
\end{equation}
where $\mu_0$ is the permeability of free space. The quantity $B_i$ is used in the present study as a derived physical feature.

When the coil field is perpendicular to the horizontal component of Earth's magnetic field $B_H$, the equilibrium deflection angle $\theta$ satisfies the tangent law
\begin{equation}
\tan\theta = \frac{B_i}{B_H}.
\label{eq:tangent}
\end{equation}
and therefore
\begin{equation}
B_H = \frac{B_i}{\tan\theta}.
\label{eq:bh}
\end{equation}
The $B_H$ values used as the ANN target are the conventional Tangent Galvanometer estimates obtained from the experimental observations. Since $\tan\theta$ and $B_i$ are also supplied to the final ANN, the model is strongly connected to the same physical relationship used to calculate the target. The ANN is therefore treated as an empirical prediction and error-modelling layer, rather than as an independent derivation of the geomagnetic field.

\section{Experimental Procedure and Dataset}
\subsection{Experimental arrangement and measurements}
The Tangent Galvanometer experiment was performed using two circular coils with 50 and 500 turns. The coil was oriented in the magnetic meridian so that the field produced at the centre was approximately perpendicular to the horizontal component of Earth's magnetic field. The current was varied systematically and the corresponding magnetic-needle deflection was recorded after the needle reached a stable equilibrium position.

For each selected current, multiple readings were taken and the average deflection angle was used for the subsequent analysis. The current direction was reversed when necessary to reduce systematic effects. The primary recorded quantities were the coil current $I$ and the average deflection angle $\theta$.

The magnetic field produced by the coil was calculated from Eq.~\eqref{eq:bi}, and the conventional estimate of the horizontal component was obtained from Eq.~\eqref{eq:bh}. Thus, each observation supplied an experimental target $B_H$ together with the measured and derived quantities used for ANN modelling.

The experimental measurements are subject to uncertainties arising from instrument resolution, current fluctuations, alignment, pivot friction, parallax, coil tolerances, and environmental magnetic disturbances. These effects produce observation-to-observation variation in the calculated $B_H$ values even though the local horizontal geomagnetic field is expected to be approximately constant.

\subsection{Data splitting and preprocessing}
The 223 retained observations were divided into training, validation, and testing subsets using a 70\%-15\%-15\% split stratified by the number of turns as shown in Table~\ref{tab:split}. Standardization of the inputs and target was fitted only to the training subset and then applied to the validation and testing subsets, thereby avoiding preprocessing leakage.

\begin{table}[H]
\centering
\caption{Dataset partitioning for ANN training, validation, and testing.}
\label{tab:split}
\begin{tabular}{lll}
\toprule
Subset & Nominal proportion & Purpose \\
\midrule
Training & 70\% & Parameter optimization and model fitting \\
Validation & 15\% & Architecture selection and early stopping \\
Testing & 15\% & Held-out final model evaluation \\
\bottomrule
\end{tabular}
\end{table}

\section{ANN Methodology}
\subsection{Input and target variables}
A single fully connected feed-forward ANN was developed using observations from both coil configurations. The number of turns was included explicitly so that one model could represent the two configurations. Table~\ref{tab:variables} lists all the five inputs to the final model such as current $I$, average deflection angle $\theta$, $\tan\theta$, induced magnetic field $B_i$, and number of turns $N$. The experimentally calculated value of $B_H$ was treated as the target which was compared with the ANN output. An important methodological point is that $\tan\theta$ and $B_i$ are not additional independent measurements. Both are derived from the information already contained in the experiment.

\begin{table}[H]
\centering
\caption{Description and role of ANN input and output variables.}
\label{tab:variables}
\begin{tabular}{lll}
\toprule
Feature & Role & Nature \\
\midrule
Current, $I$ & ANN input & Measured experimental variable \\
Average deflection, $\theta$ & ANN input & Measured experimental variable \\
$\tan\theta$ & ANN input & Derived from measured $\theta$ \\
Induced field, $B_i$ & ANN input & Derived physical field \\
No. of turns, $N$ & ANN input & Experimental variable \\
$B_H$ & ANN output & Experimentally calculated target \\
\bottomrule
\end{tabular}
\end{table}

\subsection{Experimental dataset and quality control}
The datasets corresponding to the two coils were combined. The original dataset contained 225 observations: 114 from the 50-turn coil and 111 from the 500-turn coil. Two predefined quality-control criteria were applied. First, the average deflection angle had to satisfy $0^\circ < \theta < 90^\circ$. Second, observations with a calculated target $B_H \geq 100~\mu$T were excluded. These criteria removed two observations, leaving 223 observations: 113 from the 50-turn coil and 110 from the 500-turn coil.

The summary of the retained experimental observations for both the coils is presented in Table~\ref{tab:dataset}. The 50-turn coil was operated from 12.32 to 353.70~mA, with a mean current of 84.17080~mA, whereas the 500-turn coil was operated from 1.26 to 22.35~mA, with a mean current of 8.48073~mA. The number of turns was therefore retained explicitly as an ANN input so that the combined model could distinguish the two configurations.

\begin{table}[H]
\centering
\caption{Summary of the retained experimental observations after quality-control filtering.}
\label{tab:dataset}
\begin{tabular}{lrrrr}
\toprule
Coil & No. of observations & Mean current (mA) & Minimum current (mA) & Maximum current (mA) \\
\midrule
50-turn & 113 & 84.17080 & 12.32 & 353.70 \\
500-turn & 110 & 8.48073 & 1.26 & 22.35 \\
\bottomrule
\end{tabular}
\end{table}

\subsection{ANN architecture and training}
\begin{figure}[H]
\centering
\includegraphics[width=0.95\textwidth]{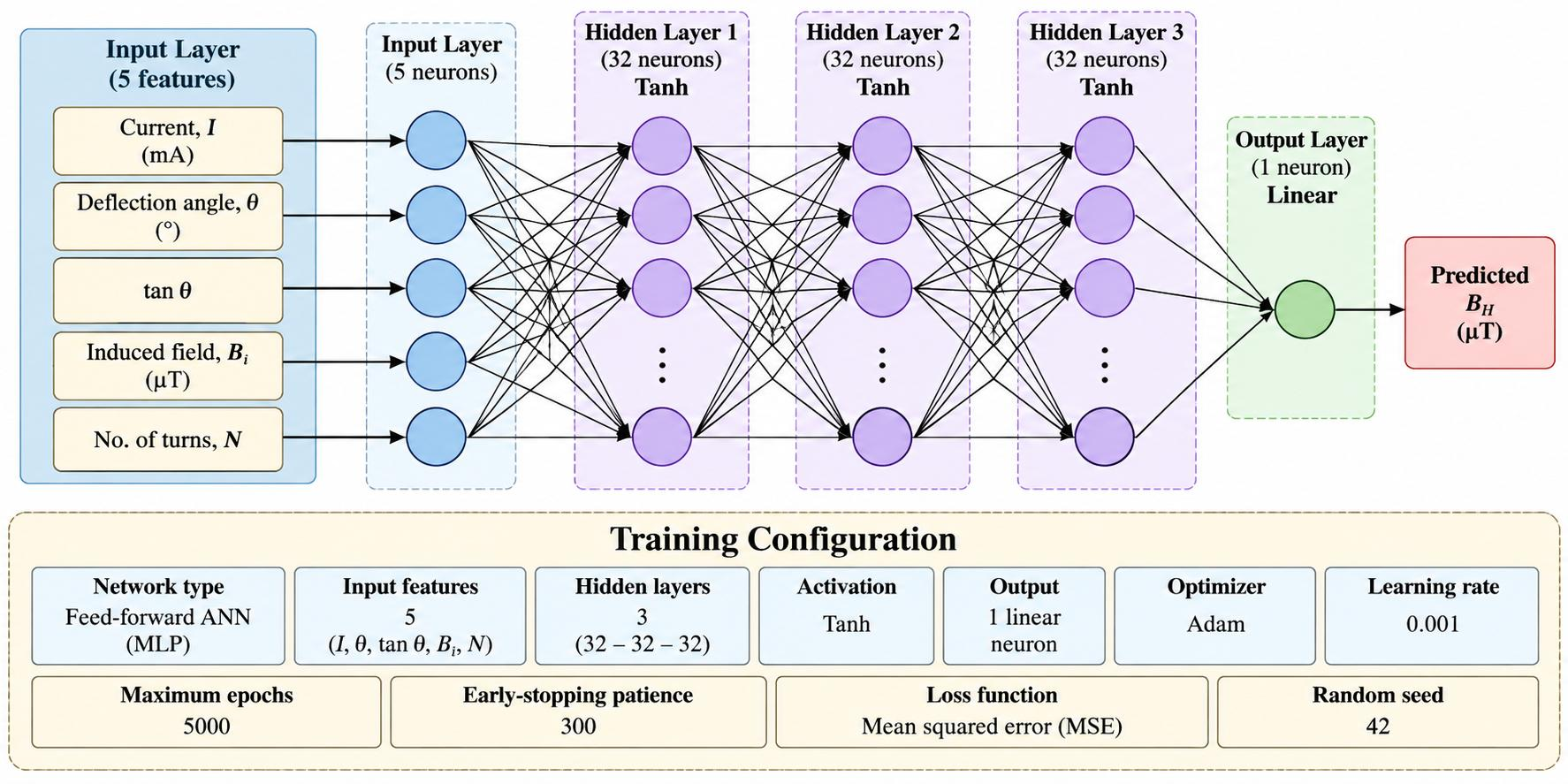}
\caption{Architecture of the feed-forward artificial neural network for $B_H$.}
\label{fig:architecture}
\end{figure}

The ANN was implemented in PyTorch. The architecture search compared Tanh and ReLU activation functions with one to five hidden layers, using 32 neurons in each hidden layer. Architecture selection was based only on the highest validation $R^2$. The test set was kept separate and was not used for model selection. The selected architecture was $5\rightarrow32\rightarrow32\rightarrow32\rightarrow1$, corresponding to 2,337 trainable parameters. This is a relatively high model capacity for a dataset containing 223 observations, so validation-based selection and early stopping were retained using the Adam optimizer with mean squared error as the loss function to avoid overfitting. Figure~\ref{fig:architecture} represents the complete feed-forward ANN architecture used to predict $B_H$.

\subsection{Evaluation metrics}
The performance of the ANN model was evaluated through various error metrics such as Coefficient of Determination ($R^2$), Mean Squared Error (MSE), Root Mean Squared Error (RMSE), Mean Absolute Error (MAE) and Mean Absolute Percentage Error (MAPE). Let $y_i$ denote the experimental target, $\hat{y}_i$ the ANN prediction, $\bar{y}$ the mean target value, and $n$ the number of observations. The residual is defined as
\begin{equation}
e_i = y_i - \hat{y}_i.
\end{equation}
The metrics are
\begin{align}
\mathrm{MAE} &= \frac{1}{n}\sum_{i=1}^{n}|y_i-\hat{y}_i|,\\
\mathrm{MSE} &= \frac{1}{n}\sum_{i=1}^{n}(y_i-\hat{y}_i)^2,\\
\mathrm{RMSE} &= \sqrt{\frac{1}{n}\sum_{i=1}^{n}(y_i-\hat{y}_i)^2},\\
R^2 &= 1-\frac{\sum_{i=1}^{n}(y_i-\hat{y}_i)^2}{\sum_{i=1}^{n}(y_i-\bar{y})^2},\\
\mathrm{MAPE} &= \frac{100}{n}\sum_{i=1}^{n}\left|\frac{y_i-\hat{y}_i}{y_i}\right|.
\end{align}
The same MAE, RMSE, and MAPE definitions were also applied to the experimental dataset. Both the ANN and experimental estimates of $B_H$ were compared with the local geomagnetic reference value of 39.0~$\mu$T at NISER, Jatni (Latitude: 20.1686$^\circ$ N, Longitude: 85.6849$^\circ$ E)~\cite{noaa}.

\section{Results and Discussions}
\subsection{Network-depth and activation-function analysis}
\begin{figure}[H]
\centering
\includegraphics[width=0.80\textwidth]{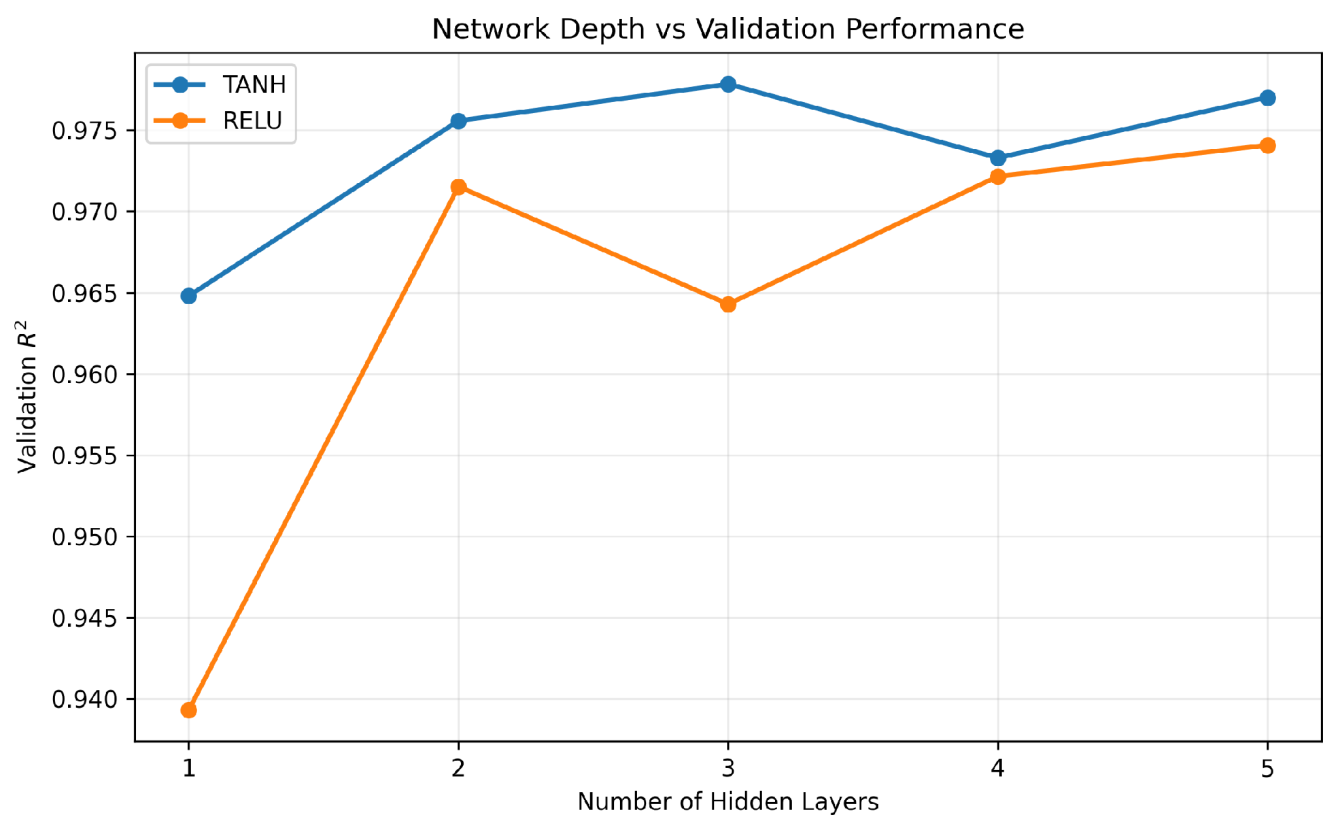}
\caption{Validation $R^2$ as a function of network depth for Tanh and ReLU activation functions.}
\label{fig:depth}
\end{figure}

The performance of the ANN architectures across activation functions and network depths was analyzed by using two activation functions (Tanh and ReLU) up to 5 hidden layers using all input variables. The model performance is summarized in Table~\ref{tab:depth}, highlighting the optimum performance for Tanh activation function with 3 hidden layers. The results also show that increasing depth does not automatically improve generalization. Figure~\ref{fig:depth} depicts the validation $R^2$ as a function of network depth for both the activation functions.

\begin{table}[H]
\centering
\caption{Performance of the ANN architectures across activation functions and network depths.}
\label{tab:depth}
\resizebox{\textwidth}{!}{%
\begin{tabular}{llrrrrr}
\toprule
Activation & Hidden layers & Best epoch & Validation $R^2$ & Test $R^2$ & Test RMSE ($\mu$T) & Test MAE ($\mu$T) \\
\midrule
Tanh & 1 & 2500 & 0.96480 & 0.98640 & 0.64800 & 0.47446 \\
Tanh & 2 & 1063 & 0.97558 & 0.98776 & 0.61471 & 0.36821 \\
Tanh & 3 & 1996 & 0.97782 & 0.99053 & 0.54076 & 0.32940 \\
Tanh & 4 & 746 & 0.97333 & 0.98939 & 0.57228 & 0.25399 \\
Tanh & 5 & 1216 & 0.97703 & 0.98789 & 0.61159 & 0.29483 \\
ReLU & 1 & 2498 & 0.93270 & 0.95460 & 1.18398 & 0.87702 \\
ReLU & 2 & 2380 & 0.97154 & 0.98124 & 0.76120 & 0.53758 \\
ReLU & 3 & 792 & 0.96427 & 0.97516 & 0.87579 & 0.59969 \\
ReLU & 4 & 1261 & 0.97239 & 0.98439 & 0.69423 & 0.44966 \\
ReLU & 5 & 668 & 0.97391 & 0.97993 & 0.78732 & 0.49440 \\
\bottomrule
\end{tabular}%
}
\end{table}

\subsection{Influence of input variables on model performance}
Three ANN models (A, B and C) were compared while keeping the network architecture, data partition, and training procedure fixed. Model A used the directly measured quantities and the coil configuration, $(I,\theta,N)$ as input variables. Model B added $\tan\theta$ as additional variable such that the input variables are $(I,\theta,\tan\theta,N)$. Model C further added the induced magnetic field $B_i$, to use four input variables $(I,\theta,\tan\theta,B_i,N)$. The purpose was to determine whether physically meaningful transformations of the measured information make the nonlinear regression problem easier for the ANN to learn. No physics-based loss term or physical constraint was included. Table~\ref{tab:models} shows the comparative study of the performance of all the three models using the test dataset. It can be seen clearly from the table that Model C estimates the optimum error metrics.

\begin{table}[H]
\centering
\caption{Test dataset performance of models A, B and C.}
\label{tab:models}
\resizebox{\textwidth}{!}{%
\begin{tabular}{lrrrrr}
\toprule
Model & Feature representation & Test $R^2$ & Test RMSE ($\mu$T) & Test MAE ($\mu$T) & Test MSE ($\mu$T$^2$) \\
\midrule
A & $I,\theta,N$ & 0.89220 & 1.82445 & 0.96491 & 3.32860 \\
B & $I,\theta,\tan\theta,N$ & 0.98721 & 0.62834 & 0.46561 & 0.39482 \\
C & $I,\theta,\tan\theta,B_i,N$ & 0.99053 & 0.54076 & 0.32940 & 0.29242 \\
\bottomrule
\end{tabular}%
}
\end{table}

\begin{figure}[H]
\centering
\includegraphics[width=0.90\textwidth]{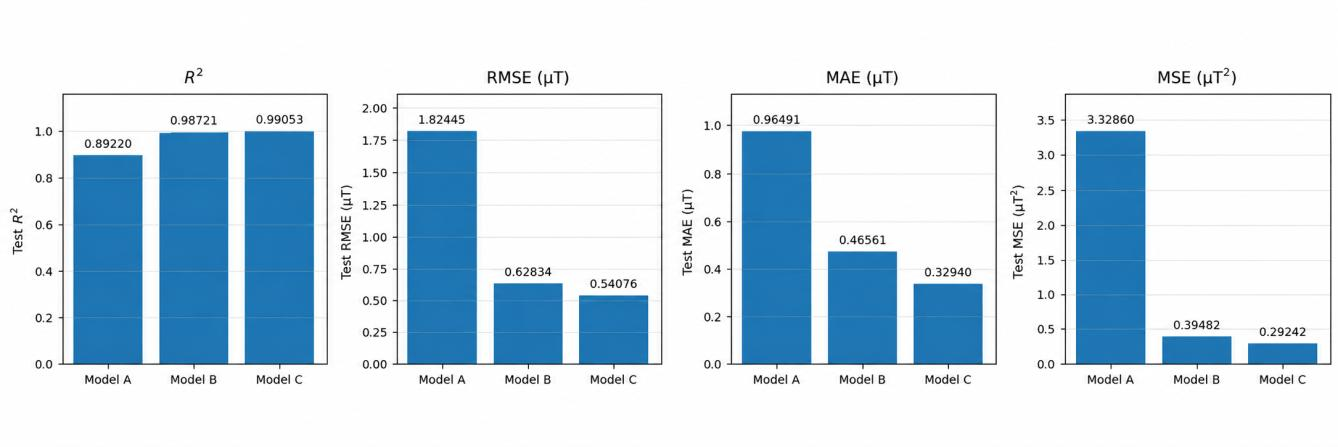}
\caption{Test dataset performance of Models A, B, and C as physically derived input variables are progressively included.}
\label{fig:modelcomparison}
\end{figure}

Figure~\ref{fig:modelcomparison} shows the performance of Models A, B, and C as physically derived input variables are progressively included. Inclusion of $\tan\theta$ in Model B provides a substantially different representation for the nonlinear regression problem as compared to Model A. Adding $B_i$ to Model C gives a further reduction in test error. It is interesting to observe that although $\tan\theta$ and $B_i$ are not independent experimental parameters, yet including these physics derived input variables improves the model performance significantly in terms of error metrices. Hence Model C was selected for further studies to estimate the value of $B_H$.

\subsection{Training and validation behavior of the final model}
\begin{figure}[H]
\centering
\includegraphics[width=0.85\textwidth]{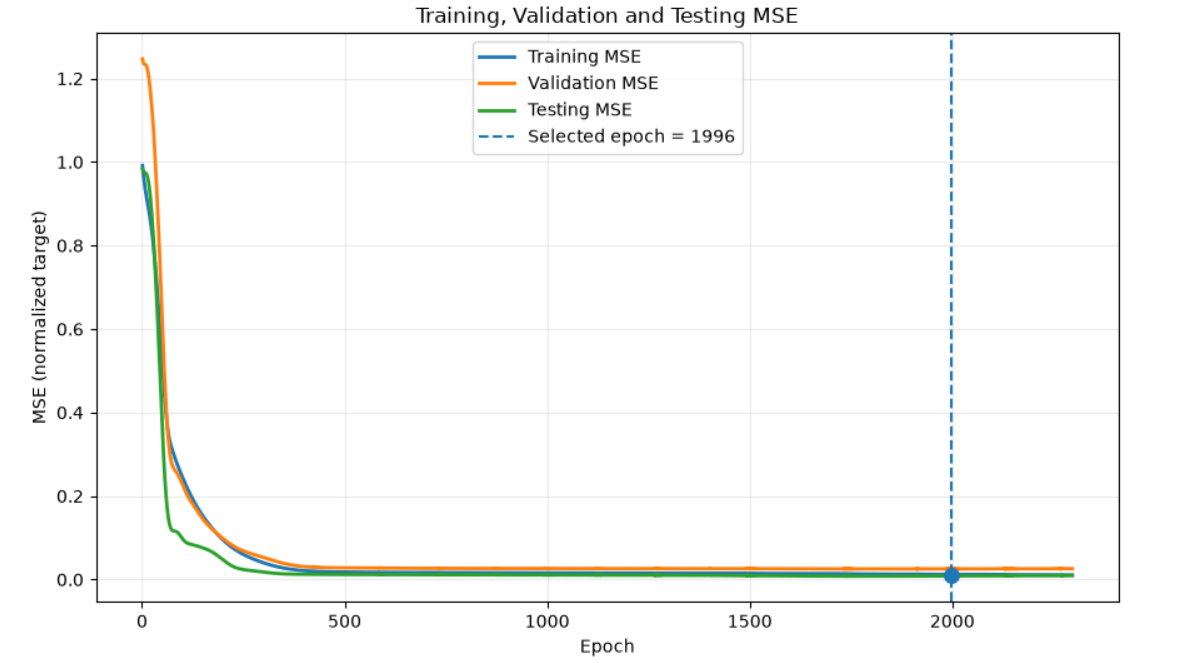}
\caption{Training and validation MSE during training of the final ANN. The dashed line marks the selected epoch (1996).}
\label{fig:training}
\end{figure}

The training and validation behavior of MSE as a function of number of epochs is shown in Figure~\ref{fig:training}. In both the cases, MSE decrease rapidly during the early epochs and then become relatively stable. The final model parameters were taken from epoch 1996, where the best validation performance was obtained. The MSE for test dataset was calculated only after the final model had been selected.

\subsection{Evaluation of performance metrics}
\begin{figure}[H]
\centering
\includegraphics[width=0.92\textwidth]{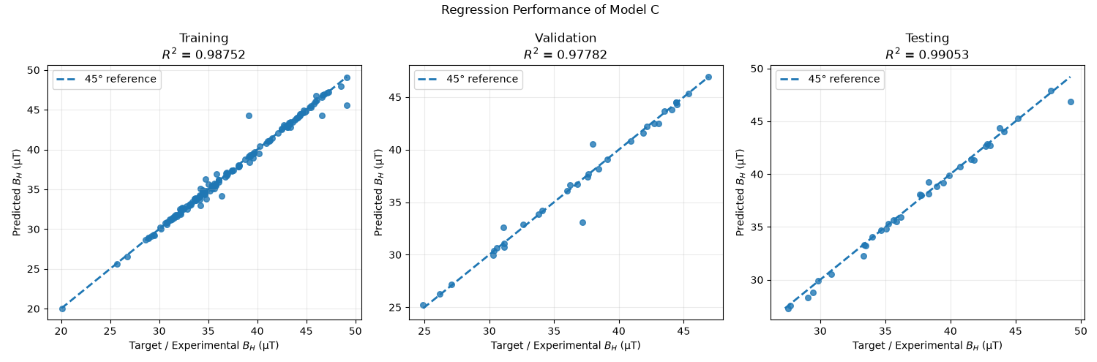}
\caption{Regression performance of Model C for the training, validation, and testing subsets; the dashed line is the 45$^\circ$ reference.}
\label{fig:regression}
\end{figure}

The regression performance of Model C for the training, validation, and test datasets is shown in Figure~\ref{fig:regression}. The $R^2$ value in each case is very close to the ideal value of 1 depicted as 45$^\circ$ reference line in the figure. The evaluation was carried out separately on the training (155), validation (34) and test (34) datasets. The test dataset was excluded from model fitting and architecture selection. The complete evaluation of the performance metrics of the final ANN model is summarized in Table~\ref{tab:metrics}.

\begin{table}[H]
\centering
\caption{Performance metrics for the combined training-validation and test dataset.}
\label{tab:metrics}
\begin{tabular}{lrrrrr}
\toprule
Dataset & $n$ & $R^2$ & RMSE ($\mu$T) & MAE ($\mu$T) & MSE ($\mu$T$^2$) \\
\midrule
Training & 155 & 0.98564 & 0.68693 & 0.28522 & 0.47187 \\
Validation & 34 & 0.97782 & 0.89941 & 0.38595 & 0.80894 \\
Testing & 34 & 0.99053 & 0.54075 & 0.32940 & 0.29241 \\
\bottomrule
\end{tabular}
\end{table}

\subsection{Residual Error analysis on test dataset}
\begin{figure}[H]
\centering
\includegraphics[width=0.80\textwidth]{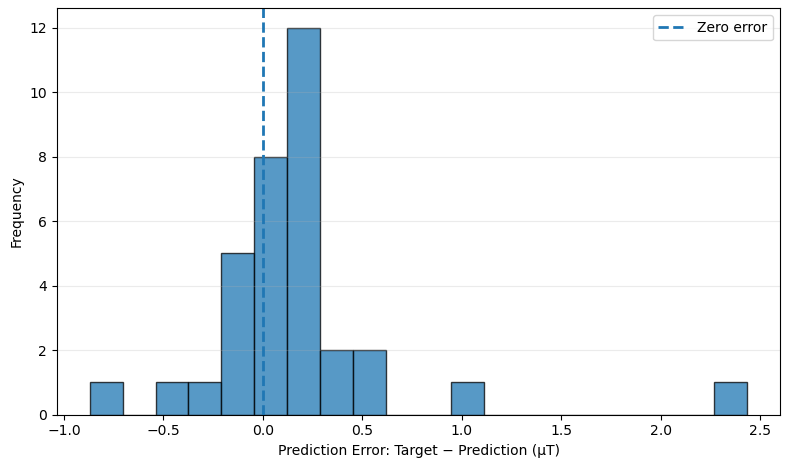}
\caption{Residual error versus observation frequency for the test dataset.}
\label{fig:residual}
\end{figure}

The residual error, defined in Eq.~(4), was analyzed for the test dataset as shown in Figure~\ref{fig:residual}. It can be observed that most of the residuals are concentrated near zero, although a few larger deviations remain, particularly on the positive side. The distribution of error indicates that the ANN reproduces most of the observations closely but does not remove the variations due to experimental measurements.

\subsection{Performance of Model C for different coil configuration}
\begin{figure}[H]
\centering
\includegraphics[width=0.82\textwidth]{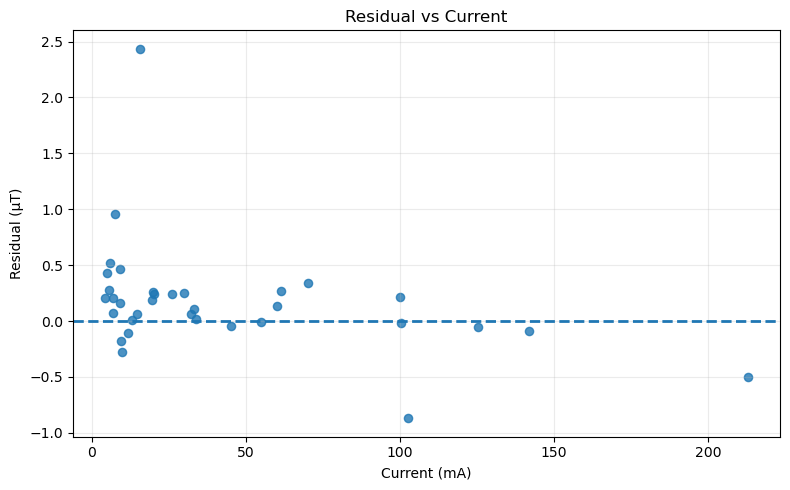}
\caption{Residuals versus applied current for the held-out test set.}
\label{fig:rescurrent}
\end{figure}

The final Model C was also examined separately for the 50- and 500-turn configurations. The 50-turn coil spans 12.32--353.70~mA, whereas the 500-turn coil spans only 1.26--22.35~mA. The model performance is shown in Table~\ref{tab:coil} through the error metrics. The larger RMSE and MAE are attributed to the substantially lower-current operating regime in the case of 500-turn configuration. Other possible contribution to larger error may come from the angular sensitivity, alignment and pivot friction in this range. These factors were not independently quantified in the present dataset, so their individual contributions cannot be separated. The residual-versus-current plot as shown in Figure~\ref{fig:rescurrent} indicates that the largest deviations occur among observations at relatively low currents as well as at a few higher-current points.

\begin{table}[H]
\centering
\caption{Performance of the combined ANN model by coil configuration.}
\label{tab:coil}
\resizebox{\textwidth}{!}{%
\begin{tabular}{lrrrrrrr}
\toprule
Coil & $N$ & Mean current (mA) & Minimum current (mA) & Maximum current (mA) & $R^2$ & RMSE ($\mu$T) & MAE ($\mu$T) \\
\midrule
50-turn & 113 & 84.17080 & 12.32 & 353.70 & 0.99678 & 0.30629 & 0.21591 \\
500-turn & 110 & 8.48073 & 1.26 & 22.35 & 0.98495 & 0.70073 & 0.44289 \\
\bottomrule
\end{tabular}%
}
\end{table}

\subsection{Comparison of Experimental and ANN predicted value of $B_H$}
\begin{figure}[H]
\centering
\includegraphics[width=0.96\textwidth]{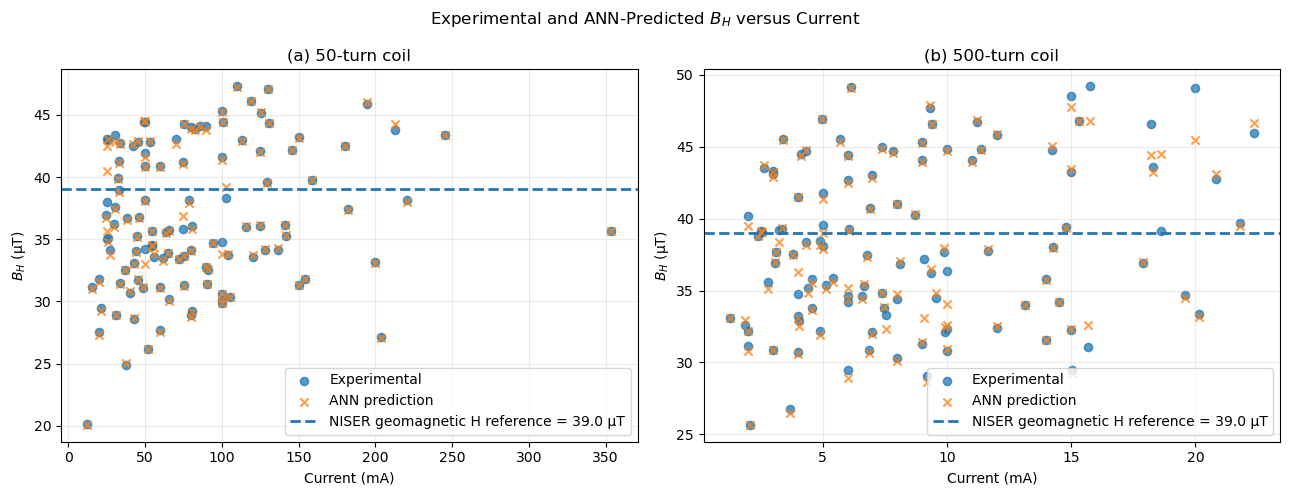}
\caption{Comparison of experimental and ANN-predicted value of $B_H$ for (a) 50-turn and (b) 500-turn coil. The blue dashed line denotes the adopted local geomagnetic horizontal reference of 39.0~$\mu$T.}
\label{fig:bhcomparison}
\end{figure}

The experimentally estimated values of $B_H$ were compared with the ANN predicted values for different coil configurations as shown in Figure~\ref{fig:bhcomparison}. For both the configurations the ANN-predicted values agree satisfactorily with most of the experimental values. The local geomagnetic reference value~\cite{noaa} is also marked by blue dashed line for comparison.

Finally, Table~\ref{tab:reference} summarizes the mean values of $B_H$ estimated from the experiment, ANN model and the error metrics relative to the local geomagnetic reference value, $B_{H,\mathrm{ref}}$. Out of all the 223 retained observations, the mean value for the experimental and ANN-predicted $B_H$ was estimated as 37.38898~$\mu$T and 37.32161~$\mu$T, respectively. The error metrics computed for both the experimental and ANN model with respect to the $B_{H,\mathrm{ref}}$ value were found to be quite similar.

\begin{table}[H]
\centering
\caption{Comparison of experimental and ANN estimates with the adopted local geomagnetic horizontal reference.}
\label{tab:reference}
\begin{tabular}{lrrrr}
\toprule
Method & Mean $B_H$ ($\mu$T) & MAE to $B_{H,\mathrm{ref}}$ ($\mu$T) & RMSE to $B_{H,\mathrm{ref}}$ ($\mu$T) & MAPE (\%) \\
\midrule
Experimental & 37.38898 & 5.09088 & 5.92924 & 13.05354 \\
ANN & 37.32161 & 5.10273 & 5.92381 & 13.08393 \\
\bottomrule
\end{tabular}
\end{table}

\section{Conclusion}
An artificial neural network was developed as a complementary data-driven model for Tangent Galvanometer measurements of the horizontal component of Earth's magnetic field ($B_H$). A combined dataset from 50-turn and 500-turn coils was screened and analyzed using a stratified training/validation/testing workflow, training-only standardization, validation-based architecture selection, and early stopping. Three different models with different numbers of input variables were compared. Addition of $\tan\theta$ produced the largest change in the prediction performance, which was improved further after including $B_i$. Model C, which used all 5 input variables, estimates the optimum error metrics, e.g. $R^2 = 0.99053$, RMSE = 0.54076~$\mu$T, and MAE = 0.32940~$\mu$T. The regression performance and residual error analysis of Model C was evaluated for optimization. The performance of Model C was also examined separately for the different coil configurations, i.e., the 50-turn and 500-turn Tangent Galvanometer coils. The comparison provides an additional assessment of the ability of the ANN model to represent the experimental data obtained from different coil configurations and demonstrates the applicability of the model across the two experimental conditions.

Finally, the experimental and ANN predicted values of $B_H$ were compared with the adopted local geomagnetic-reference (39.0~$\mu$T). The mean experimental value of $B_H$ was found to be 37.38898~$\mu$T, while the corresponding mean value predicted by the ANN was 37.32161~$\mu$T. The experimental and ANN-based estimates show very similar deviations from the adopted geomagnetic reference value. This indicates that the ANN model closely represents the experimental measurements, rather than introducing a substantial shift in the estimated value of $B_H$.

Overall, this study shows that ANN modelling can be used with a standard undergraduate Tangent Galvanometer experiment to study experimental variations and nonlinear relationships while keeping the connection with the underlying physics. Further work with larger datasets and repeated experiments is needed to check the reliability of the model. Including factors such as the surrounding magnetic field, temperature, humidity, instrument alignment, and coil geometry may further improve the prediction. Direct local geomagnetic measurements and better uncertainty analysis may also help the ANN predictions agree more closely with the actual local geomagnetic field.

\end{document}